\documentclass[prl,twocolumn]{revtex4}
\usepackage{amssymb}\usepackage{amsmath}\usepackage{graphicx,epsfig}
\usepackage{mathrsfs} 
\usepackage{amsbsy,latexsym}
\usepackage[matrix,frame,arrow]{xypic}\input{Qcircuit}
\usepackage{epsfig,eepic,epic}
\usepackage[hypertex,linkcolor=red]{hyperref}

\def\>{\rangle} \def\<{\langle}
\newcommand{\Tr}{\operatorname{Tr}}

\def\Choi#1{\operatorname{Choi}(\map{#1})}\def\Chov#1{\operatorname{Choi}^{-1}({#1})}

\def\vec#1{{\boldsymbol{#1}}}
\def\map#1{{\mathscr{#1}}}  \def\sH{\mathcal{H}}
\def\i{{\rm in}}\def\o{{\rm out}}\def\o{{\rm out}}
\def\set#1{{\sf #1}}\def\sJ{\set{J}}\def\sH{\set{H}}\def\sT{\set{T}}
\def\Bnd#1{\set{Lin(#1)}}\def\Bndd#1{\set{Lin}(#1)}
\def\UStt#1{\set{Lin}(#1)}
\newtheorem{Rule}{Rule}
\newtheorem{theorem}{Theorem}
\renewcommand{\geq}{\geqslant}\renewcommand{\leq}{\leqslant}
\begin{document}
\title{Quantum Circuits Architecture}

\author{G. Chiribella} 
\affiliation{QUIT Group,
  Dipartimento di Fisica ``A. Volta'', via Bassi 6, I-27100 Pavia,
  Italy and CNISM.} 

\author{G. M. D'Ariano} 
\affiliation{QUIT Group,
  Dipartimento di Fisica ``A. Volta'', via Bassi 6, I-27100 Pavia,
  Italy and CNISM.} 

\author{P. Perinotti} 
\affiliation{QUIT Group,
  Dipartimento di Fisica ``A. Volta'', via Bassi 6, I-27100 Pavia,
  Italy and CNISM.} 

\begin{abstract}
  We present a method for optimizing quantum circuits architecture.
  The method is based on the notion of {\em quantum comb}, which
  describes a circuit board in which one can insert variable
  subcircuits. The method allows one to efficiently address novel
  kinds of quantum information processing tasks, such as
  storing-retrieving, and cloning of channels.
\end{abstract}

\maketitle 

Quantum Mechanics plays a crucial role in the technology of high
precisions and high sensitivities, e.~g. in frequency
standards \cite{FRST}, quantum lithography \cite{SCUL}, two-photon
microscopy \cite{QLIT}, clock synchronization \cite{QPOS}, and
reference-frame transfer \cite{QCOR}.  In all these applications the
essential problem is to achieve very high precision in: {\em i)}
determining parameters; {\em ii)} executing a transformation that
depends on unknown parameters.  Since the parameters are generally
encoded with a transformation \cite{note}, as in the whole class of
\emph{quantum metrology} problems \cite{metro}, and since the estimation
itself can be considered as a special case of transformation (with
classical $c$-number output), both tasks {\em i)} and {\em ii)} can be
reduced to the general problem of {\em executing a desired
  transformation depending on an unknown transformation}. Taking into
account the possibility of exploiting $N$ uses of the unknown
transformation, the problem becomes to build a quantum circuit that has $N$
circuits as input, and achieves the desired transformation as an
output. This is what we call a {\em quantum circuit board}.

\begin{figure}[h]
\epsfig{file=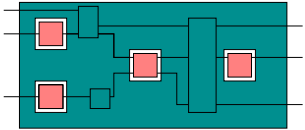,width=4cm}
\caption{A quantum circuit board.}\label{fig:board}
\end{figure}

A quantum circuit board is a network of gates in which there are $N$ slots with open ports for the insertion of
$N$ variable sub-circuits (see Fig. \ref{fig:board}).
Since generally it is impossible even in principle to achieve the
desired transformation exactly, the main task here is to optimize the
circuit board according to a given figure of merit---e.~g. the
gate-fidelity with the target transformation. A typical
example is the optimal {\em cloning of an undisclosed transformation}
$U$, which will be operated by a board with $N$ slotted uses of $U$,
and achieving overall in-out unitary transformation which is the
closest possible to $U^{\otimes M}$ with $M>N$. We emphasize that
generally the overall in-out transformation of the board and the
slotted ones can be of any kind, including measurements and
state-preparations (i.~e.  transformations to and from $c$-numbers),
and that in some specific situations the slotted transformations can
be even different from each other.

In previous literature some special cases of circuit-board
optimization have been considered, e.~g.  in regards to the
discrimination of two unitary transformations with $N$ uses in
parallel \cite{lop}, or in sequence \cite{duan}, and to the
problem of phase estimation \cite{metro}, where an optimal quantum circuit
architecture has been determined \cite{mosca}.  However, the optimal
architecture for estimating arbitrary unitaries is still unknown, and
no systematic method for the general problem of circuit board
optimization is available yet.

In this Letter we present a complete method for optimizing the
architecture of quantum circuit boards. After providing a convenient
description of circuit connectivity, we introduce the notion of
\emph{quantum comb}, which describes all possible transformations
operated by a quantum circuits board, and generalizes the notion of
quantum channel to the case where the inputs are quantum circuits,
rather than quantum states. We then present the optimization method,
based on the convex structure of the set of quantum combs.  The method
allows one to reduce the apparently untractable problem of optimal
circuit architecture to the optimization of a single positive operator
with linear constraints.  Finally, we give two new applications in
which the present approach dramatically simplifies the solution of the
problem.

A quantum circuit operates a transformation from input to output,
and is graphically represented by a box with input and output wires
symbolizing the respective quantum systems. The quantum systems of
different wires are generally different, and may also vary from input
to output.  Let us associate Hilbert spaces $\sH_\i$ ($\sH_\o$) to all
input (output) wires, and denote by $\rho_\i$ ($\rho_\o$) the
corresponding states.  The action of the circuit is generally
probabilistic, i.~e.  different in-out transformations can randomly
occur, as in a measurement process.  Each transformation is described
by a linear map $\rho_\i\to \map{C} (\rho_\i) = k \rho_\o$, with the
proportionality factor $0\leq k=\Tr[\map{C}(\rho_\i)]\leq 1$ giving
the probability that $\map{C}$ occurs on state $\rho_\i$.  To describe
a legitimate quantum transformation, the map
$\map{C}:\UStt{\sH_\i}\to\UStt{\sH_\o}$ \cite{noted} has to be
completely positive (CP) \cite{noteCP} and trace non-increasing.
Trace-preserving maps---i.e. deterministic
transformations---are called \emph{quantum channels}. Notice
that a map $\map{C}$, rather than representing a specific circuit, is
univocally associated to the equivalence class of all circuits
performing the same in-out transformation.

The linear map $\map{C}$ can be conveniently rewritten using the so-called
"Choi-Jamio\l kowski" representation \cite{Zyckowski}, corresponding
to the following one-to-one correspondence between linear maps
$\map{C}:\UStt{\sH_\i}\to\UStt{\sH_\o}$ and linear operators
$C\in\Bndd{\sH_\o\otimes\sH_\i}$ given by
\begin{eqnarray}\label{Choi}
C&=&\Choi{\map{C}}:=\map{C}\otimes\map{I}(|\Omega\>\<\Omega|).\\
\label{Choi-1}
\map{C}(\rho)&=&\Chov{C}(\rho):=\Tr_{\i} [(I_{\o}\otimes\rho^T ) C],
\end{eqnarray}
where $\map I$ is the identity map, $|\Omega\>$ is the unnormalized
maximally entangled state
$|\Omega\>=\sum_n|n\>|n\>\in\sH_\i^{\otimes 2}$, and $T$
denotes transposition with respect to the orthonormal basis $\{|n\>\}$
for $\sH_\i$.  The map $\map C$ is CP if and only if the operator
$C$---called {\em Choi operator}---is positive \cite{Choi}.

Two quantum circuits can be connected in all the ways allowed by the
physical matchings between input and output wires (see e.~g. Fig.
\ref{f:connect}, where the wires labelled $\vec{d}$ are connected):
a connection will result in the composition of the corresponding CP
maps, and hence of the corresponding Choi operators.  Since building a
quantum network means connecting many circuits, it is crucial to have
a handy way to describe circuit connectivity with minimum overhead of
notation.  We provide here three simple rules that accomplish this
goal:
\begin{figure}[htb]
\begin{center}
$$
\begin{matrix}
\begin{matrix}
\Qcircuit @C=1em @R=.7em @! R {
\rstick{a} & &\multigate{1}{\map{A}} & \qw &\lstick{c} \\ 
\rstick{b} & & \ghost{\map{A}} &\qw & \lstick{\vec d}}
\end{matrix},\quad
\begin{matrix}
\Qcircuit @C=1em @R=.7em @! R {
\rstick{\vec d} & &\multigate{1}{\map{B}} & \qw &\lstick{f} \\ 
\rstick{e} & & \ghost{\map{B}} &\qw & \lstick{g}}
\end{matrix}
\end{matrix}
$$
$$
\begin{matrix}
\Qcircuit @C=1em @R=.7em @! R {
\rstick{a} && \multigate{1}{\map{A}} & \qw &\qw &\qw&\lstick{c} \\ 
\rstick{b} &&\ghost{\map{A}} &\qw\vec d & \multigate{1}{\map{B}} & \qw &\lstick{f} \\
\rstick{e} && \qw & \qw & \ghost{\map{B}} &\qw&\lstick{g} }
\end{matrix}\equiv
\begin{matrix}
\Qcircuit @C=1em @R=.7em @! R {
\rstick{a} && \multigate{2}{\map{C}} & \qw&\lstick{c} \\ 
\rstick{b} &&\ghost{\map{A}} & \qw &\lstick{f} \\
\rstick{e} &&\ghost{\map{A}} & \qw &\lstick{g} }
\end{matrix}
$$
\smallskip
$$\quad A=\Choi{A},\,B=\Choi{B},\, C=\Choi{C}=A*B$$
\caption{Connection of two quantum circuits $\map A$ and $\map B$. Wires are labelled according to Rule
  \ref{labelling}. The Choi operator the resulting circuit $\map C$ is given by the link product of Rule \ref{comprule}.\label{f:connect}}
\end{center}
\end{figure}

\begin{Rule}[Labelling]\label{labelling} Each quantum wire is marked with a different label, except for wires that are connected, which are identified with the same
  label.
\end{Rule}


\begin{Rule}[Multiplication]\label{multrule} The multiplication of two Choi operators
  $A\in\Bndd{\sH_{a,b,c,\vec{d}}}$ and
  $B\in\Bndd{\sH_{\vec{d},e,f,g}}$ is intended in the tensor fashion, i.e.   $A B=(A \otimes I_{e,f,g})(I_{a,b,c} \otimes B)$.
\end{Rule}
\begin{Rule}[Composition]\label{comprule} The connection of two circuits with Choi
  operators $A$ and $B$---acting on Hilbert spaces labelled according
  to Rule \ref{labelling}---yields a new circuit with Choi operator $C$
  given by the {\em link product}
\begin{equation}\label{link-prod-def}
C=A*B =\Tr_\sJ[A^{\theta_\sJ} B],
\end{equation}
$\theta_\sJ$ denoting partial transposition over the Hilbert space $\sJ$  of
the connected wires, and the multiplication in square brackets following Rule \ref{multrule}.
\end{Rule}
Rule \ref{comprule} directly follows from Eqs. (\ref{Choi}) and
(\ref{Choi-1}).  Notice that due to invariance of trace under cyclic
permutations, the link product is commutative: $A * B = B * A$.
Using it, the action of a linear map $\map C$ on a state
$\rho$ in Eq. (\ref{Choi-1}) can be rewritten as $\map C (\rho)
= C * \rho$.  Assembling many circuits $\map C_1, \map C_2, \dots,
\map C_k$ yields a quantum network whose Choi operator is simply given
by $C =C_1 * C_2 * \dots * C_k$.



\begin{figure}[h]
\epsfig{file=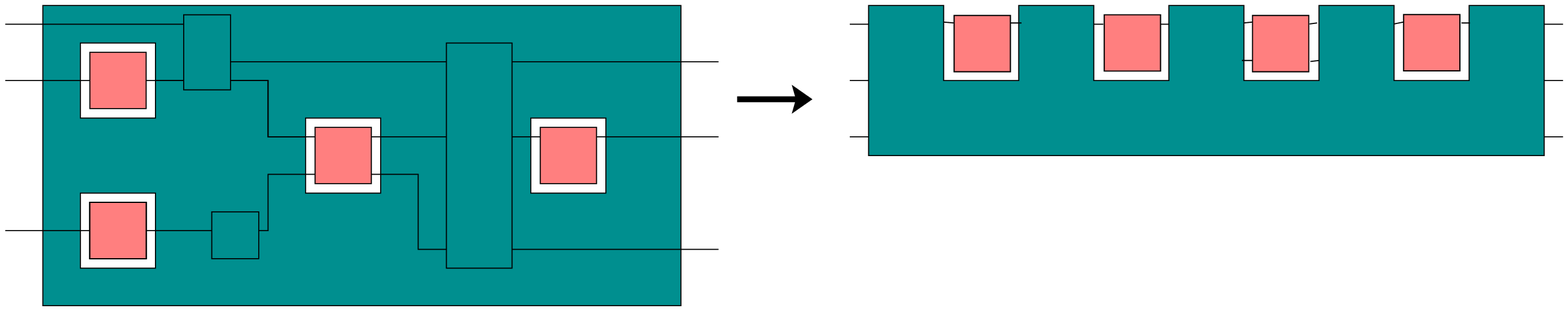,width=6cm}
\vskip .5cm
\epsfig{file=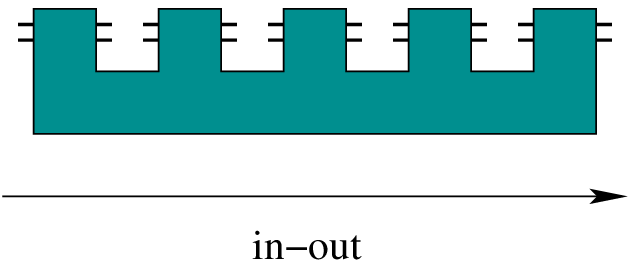,width=5cm}
\caption{Every circuit-board can be reshaped in form of a "comb", with
  an ordered sequence of slots, each between two successive teeth. The
  pins represent quantum systems, entering or exiting from the board
  (the horizontal arrow represents the quantum information flow).}\label{fig:arrow}
\end{figure}

We are now ready to treat quantum circuit boards. To start with, we
consider the case of a \emph{deterministic} circuit board, i.e. a
network of quantum channels with $N$ open slots for the insertion of
variable subcircuits.  It is clear that by reshuffling and stretching
the internal wires any circuit board can be reshaped in the form of a
"comb", with an ordered sequence of slots, each between two successive
teeth, as in Fig.  \ref{fig:arrow}. The order of the slots is the
causal order induced by the flow of quantum information in the circuit
board.  We label the input systems (entering the board) with even
numbers $2 n$, and the corresponding output systems (exiting the
board) with odd numbers $2 n+1$, with $n$ ranging from 0 to $N$.

\begin{figure}[ht] \epsfig{file=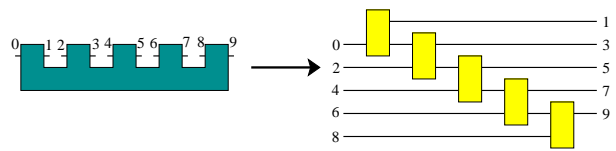,width=8cm}
\caption{Each quantum comb is equivalent to a causal network, with all
  inputs on the left and all outputs on the right. The Choi operator
  of a comb is the Choi operator of the corresponding causal network.}
\label{fig:general}
\end{figure}

A quantum comb with $N$ slots is clearly equivalent to a concatenation
of $N+1$ \emph{channels with memory},
which is in turn equivalent to \emph{causal network}, namely a network
where the quantum state of the output systems up to time $n$ does not
depend on the state of the input systems at later times $n' > n$, with
$n,n' \in \{0, 1, \dots, N\}$  \cite{wernermem}. The causal network can be easily
obtained by redrawing the comb as an equivalent circuit with all
inputs on the left and all outputs on the right, as in Fig.
\ref{fig:general}.
We define the Choi operator of a quantum comb as the Choi operator $R$ of
the corresponding causal network.  In
terms of the Choi operator $R$, causality is equivalent to a set of linear constraints 
\begin{equation}\label{causaln}
\begin{split}
\Tr_{2n+1}\left[R^{(n)}\right]&=I_{2n}\otimes R^{(n-1)}, \quad n=0,\ldots N,\\
R^{(N)}&\equiv R,\quad R^{(-1)}=1,
\end{split}
\end{equation}
where $\Tr_{2n+1}$ denotes the partial trace over the Hilbert space $\sH_{2n+1}$ of the output wire
labeled $2n+1$, $I_{2n}$ the identity operator over the Hilbert space $\sH_{2n}$ of the input wire
labeled $2n$, $R^{(n)}=\operatorname{Choi}(\map{C}^{(n)})$, and $\map{C}^{(n)}$ is the map of the
$(n+1)$-{\em subnetwork} from the first $n+1$ inputs to the first $n+1$ outputs. 
Precisely, we have the following 
\begin{theorem}\label{ChoiCharacterization} Every positive operator $0\leq R\in\Bndd{\otimes_{j=0}^{2N+1}\sH_j}$ satisfying the linear constraints
 (\ref{causaln}), is the Choi operator of a deterministic quantum comb.
\end{theorem} {\bf Proof} By definition, it is enough to show that any
operator $R\geq 0$ normalized as in Eq. (\ref{causaln}) is the Choi
operator of a causal network.  A causal network with $N+1$
input/output pairs is described by a family of channels $\map
C^{(n)}$, $n=0, 1, \dots ,N$ with the property $\Tr_{2n+1} [ \map
C^{(n)} (\rho^{(n)})] = \map C^{(n-1)} \left(\Tr_{2n} [\rho^{(n)}]
\right)$, for any state $\rho^{(n)}$ of the first $n+1$ input systems.
Using the correspondence of Eq.  (\ref{Choi-1}), one can easily see
that this is equivalent to the normalization of Eq.  (\ref{causaln}).
\medskip

\begin{figure}[ht]\bigskip
\centerline{(a)\quad\epsfig{file=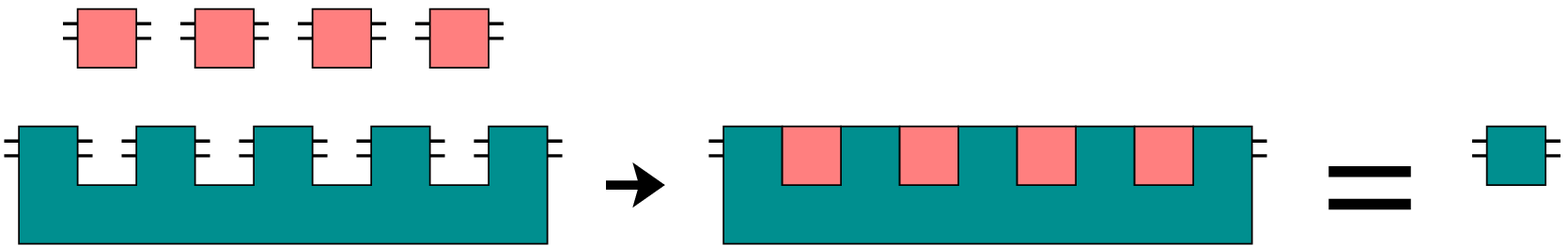,width=6cm}}\smallskip
\centerline{(b)\quad\epsfig{file=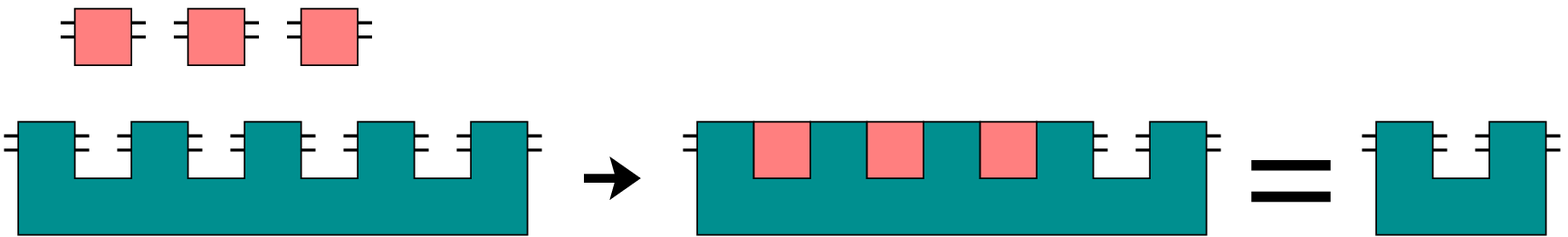,width=6cm}}\smallskip
\centerline{(c)\quad\epsfig{file=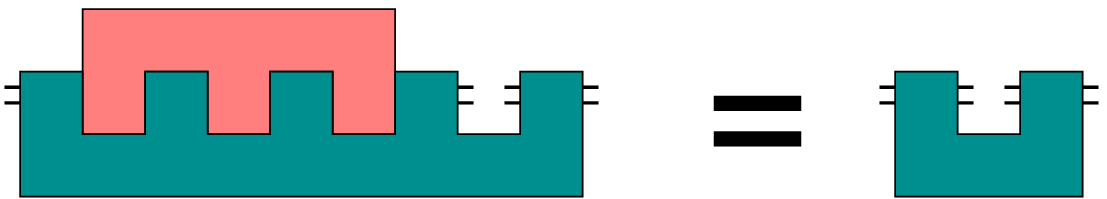,width=6cm}}\smallskip
\caption{A quantum comb realizes different transformations of quantum
  circuits, namely it can send (a) A series of channels into a
  channel.  (b) A series of channels into a comb, or (c) An input
  comb into an output comb.}\label{fig:combs}
\end{figure}

A quantum comb transforms a series of $N$ input circuits $\map C_1, \dots , \map C_N$  into an output circuit $\map C'$ depending on them (Fig. \ref{fig:combs}a). 
This transformation of circuits corresponds to an $N$-linear CP-map
that sends the input Choi operators into the ouput Choi operator according to $C' = C_1 * \dots * C_N * R$, with $R$ the Choi
operator of the comb.  
We call the mapping between circuits $\{\map C_1, \dots, \map C_N\}
\mapsto \map C'$ \emph{supermap} as it sends channels into channels,
rather than states into states.
Notice that, depending on the number of slots that are saturated a
quantum comb can transform a series of circuits into a comb (Fig.
\ref{fig:combs}b), or, more generally, a comb into a comb (Fig.
\ref{fig:combs}c).  As a matter of fact, a quantum comb realizes many
possible mappings, all obtained by the link product with its Choi
operator $R$. Therefore, the quantum comb can be completely identified
with its Choi operator. Remarkably, also the converse is true: any
abstract supermap sending channels into channels in a CP fashion can
be physically realized by a quantum comb \cite{noteSuper}.

The tools presented above provide a powerful method for optimizing
quantum circuit architecture. Suppose we want to design a circuit
board maximizing some convex figure of merit, e.g. the fidelity of the ouput
circuit $\map C'$ with a desired unitary gate $\map U$. In our
framework the optimization of the board architecture is reduced to the
search of the optimal operator $R \geq 0$ with the linear constraints
(\ref{causaln}).  This is a standard problem of \emph{convex
  optimization}, for which efficient algorithms are known.  Basically,
we only need to implement the search on the extremal points of the convex
set of Choi operators.  Moreover, the complexity of the search can be
dramatically reduced by exploiting additional constraints, e.g.
symmetry properties of the circuit board.  The optimal Choi operator
will finally single out the optimal architecture, automatically
deciding if the $N$ slots of the circuit board have to be connected in
a causal order or in parallel, or in any combination of the two.

We illustrate our method in two concrete applications.  The first
application is the {\em optimal universal cloning of unitary
  transformations}, i.~e.  the problem of designing a quantum board
that optimally achieves the $N\to M$ cloning of an unknown unitary
$U\in\mathbb{SU}(d)$ in dimension $d$. The board has $N$ slots
containing $N$ identical uses of the unknown unitary $U$ and performs
a transformation which is the closest possible to $U^{\otimes M}$.
\begin{figure}[ht]
\epsfig{file=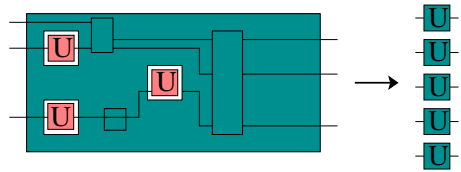,width=6cm}
\caption{Cloning of unitary transformations. A quantum board with
  $N=3$ input slots, designed to optimally emulate $M=5$ uses of the
  unknown unitary transformation $U$ with maximum channel fidelity.}
\label{fig:cloningu}
\end{figure}
Using channel fidelity as the figure of merit,
the problem is to find the Choi operator $R$ that maximizes the
average over all unitaries of the overlap $F_U (N,M) =1/(d^{2M +N})  \<
U|^{\otimes M} \< U^*|^{\otimes N}~ R~ |U\>^{\otimes M}
|U^*\>^{\otimes N}$, where $|U\> =  (U \otimes I ) |\Omega\>$, with
$|\Omega\> = \sum_n |n\>|n\>$.  
The architecture optimization is then reduced to a standard convex analysis problem.  For $N= 1$ and
$M=2$, we derived the optimal quantum board, achieving fidelity $F^{clon}(1,2) = (d +
\sqrt{d^2-1})/d^3$, significantly higher than the classical threshold reached by the optimal
estimation of a unitary $F^{est} (1,2)= 6/d^4$ for $d>2$, $F^{est} (1,2)=5/16$ for qubits
\cite{InPrep}, this showing the advantage of coherent quantum information processing over any
classical cloning strategy.

\begin{figure}[h]
\epsfig{file=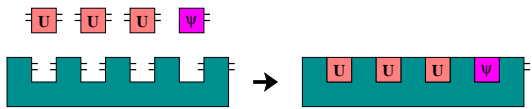,width=7cm}
\caption{Quantum-algorithm learning. One wants to run an undisclosed unitary $U$
  on a quantum state $\psi$ which is available after the lapse of time in which the uses of $U$
  are available.}
\label{fig:learning}
\end{figure}

Another interesting application is the storage and retrieval of an
undisclosed unitary transformation $U$ from $N$ uses, also called
\emph{optimal quantum-algorithm learning}. The problem arises from the
need of running an undisclosed algorithm (available for $N$ uses) on
an input state $\psi$ which will be available at later time. To this
purpose one can slot the $N$ uses of $U$ in a quantum circuit board,
put the output state of the board in a quantum memory, and, when the
input state will be available, use the memory to recover the unitary.
The series storing-retrieving is represented by the quantum comb in
Fig.  \ref{fig:learning}, which can be cut into two parts, a storing
one including only the uses of $U$, and a retrieving one including
$\psi$ (the output state of the first part is stored in a quantum
memory and  is then fed in the second part).
Also in this case our method reduces the optimization to a convex analysis problem.  For $N=1$ and
$N=2$ we found average gate-fidelities $F=\frac{2}{d^2}$ and $F=\frac{3}{d^2}$, respectively, which
coincide with the value attained by the optimal estimation of unitaries \cite{InPrep2}. Remarkably,
the optimal universal storing/retrieving of unitaries does not need a coherent interaction with the
quantum memory at the retrieving stage---which is purely classical---, but only an entangled input
state at the storing stage.  This means that the quantum memory is not really needed, and that the
learned algorithm can be executed an unlimited number of times with constant performance. The
situation is radically different if no entanglement is allowed in the storing stage: in this case
the optimal retrieving is purely quantum, yielding an optimal learning that is forgetful.



\begin{figure}[h]
\epsfig{file=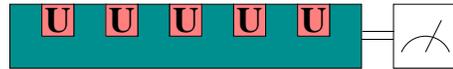,width=6cm}
\caption{Quantum comb for the estimation of unitary transformations
  with multiple uses.}
\label{fig:estimator}
\end{figure}

We conclude by mentioning the extension of our method to the
optimization of probabilistic circuit boards, containing measuring
devices that produce different transformations depending on random
outcomes. The probabilistic comb corresponding to outcome $i$ will
have Choi operator $R_i$, with the sum over all outcomes $\sum_i
R_i=R$ giving the Choi operator of a deterministic comb. Indeed,
introducing a classical register with orthogonal states $|i\>\in
\sH_C$ we can define $\widetilde R = \sum_i R_i \otimes |i\>\< i|$,
which is the Choi operator of a deterministic comb with
$\widetilde{\sH}_{2N+1}:=\sH_{2N+1} \otimes \sH_C$. The comb
corresponding to $R_i$ is then obtained after applying the comb of
$\widetilde R$, by measuring the register on the basis $\{|i\>\}$ and
postselecting outcome $i$. Probabilistic combs are a fundamental tool
to address the optimized circuit architecture for estimating unknown
transformations with multiple copies (see Fig.  \ref{fig:estimator}).
Again, by optimizing the operators $\{R_i\}$ one will automatically
determine the optimal disposition of the unitaries in the circuit, a
problem whose solution is up to now known only in the very special
case of phase estimation \cite{mosca}.

\acknowledgments This work has been supported by Ministero Italiano dell'Universit\`a e della Ricerca
(MIUR) through PRIN 2005.


\begin{thebibliography}{99}
\bibitem{note} This is the case of frequency standards (where one
  needs to estimate a phase-shift of a two-level oscillator), or of
  clock synchronization. The encoding information into parametrized
  unitary transformations is also exploited in quantum information
  tasks, e.g. in some quantum cryptographic protocols and quantum
  algorithms.
\bibitem{noted} Here we consider only finite dimensions, and denote
  the linear space of operators on $\sH$ as $\Bnd{H}$.
\bibitem{noteCP} The map $\map{C}$ has to preserve positivity of
  quantum states even when acting on a part of a compound system.
\bibitem{noteSuper} This can be proved within an axiomatic
  introduction of combs and supermaps, where a realization theorem
  holds stating that any CP supermap from combs to combs has a
  physical scheme corresponding to another comb.
\end{thebibliography}
\end{document}